# Changes in the morphology of interstellar ice analogues after hydrogen atom exposure

Mario Accolla,[ab†] Emanuele Congiu,[*a] François Dulieu,[a] Giulio Manicò,[b] Henda Chaabouni,[a] Elie Matar,[a] Hakima Mokrane,[a] Jean Louis Lemaire[a] and Valerio Pirronello[b]

The morphology of water ice in the interstellar medium is still an open question. Although accretion of gaseous water could not be the only possible origin of the observed icy mantles covering dust grains in cold molecular clouds, it is well known that water accreted from the gas phase on surfaces kept at 10 K forms ice films that exhibit a very high porosity. It is also known that in the dark clouds $H_2$ formation occurs on the icy surface of dust grains and that part of the energy (4.48 eV) released when adsorbed atoms react to form $H_2$ is deposited in the ice. The experimental study described in the present work focuses on how relevant changes of the ice morphology result from atomic hydrogen exposure and subsequent recombination. Using the temperature-programmed desorption (TPD) technique and a method of inversion analysis of TPD spectra, we show that there is an exponential decrease in the porosity of the amorphous water ice sample following D-atom irradiation. This decrease is inversely proportional to the thickness of the ice and has a value of $\phi_0 = 2 \times 10^{16}$ D-atoms cm$^{-2}$ per layer of $H_2O$. We also use a model which confirms that the binding sites on the porous ice are destroyed regardless of their energy depth, and that the reduction of the porosity corresponds in fact to a reduction of the effective area. This reduction appears to be compatible with the fraction of $D_2$ formation energy transferred to the porous ice network. Under interstellar conditions, this effect is likely to be efficient and, together with other compaction processes, provides a good argument to believe that interstellar ice is amorphous and non-porous.

## 1. Introduction

Amorphous solid water (ASW) is reputed to be the most abundant form of water in the Universe, thanks to its propensity for forming, molecule after molecule, as a deposit on interstellar dust particles.[1,2] The accretion of icy mantles (predominantly constituted by ASW) on silicate and carbonaceous dust grains takes place in interstellar molecular clouds. Some of them are cold (~10 K) and dense ($10^4$ - $10^6$ cm$^{-3}$) regions where gaseous species heavier than hydrogen and helium freeze out onto the grains. Beside $H_2O$ and $CO$, other species constitute the inventory of icy mantles such as $CO_2$, $CH_3OH$, $CH_4$, $H_2CO$ and $NH_3$,[3,4] as a result of the gas-grain interactions occurring on the surface of interstellar dust. The most important and abundant molecular species as well, $H_2$, is formed likewise thanks to reactions occurring on amorphous water ice in molecular clouds.[5,6] In addition, a significant fraction of the gas phase species is constituted by atomic hydrogen whose abundance is mostly governed by the destruction of $H_2$ due to cosmic rays. It has been evaluated that, in dense clouds, the average number density ratio of atomic hydrogen component compared to the molecular one is ~ 0.1%.[7] In a dark cloud, atomic hydrogen thus represents the third most abundant gas-phase species, after $H_2$ and He.

Along with atoms and molecules available in the gas-phase and the condensed/adsorbed species on the surface of dust grains, the roughness of the icy grains can also greatly affect the efficiency of surface reactions. In fact, the porosity of interstellar ices governs both the efficiency of $H_2$ formation and how the formation energy is distributed between the reaction site and the internal and translational energy in the molecule.[8] For instance, it has been recently demonstrated that the excess energy of this reaction is mostly deposited in the ice mantle under dense cloud conditions of the interstellar medium.[9] It was also suggested that this exothermic reaction can induce the desorption of already adsorbed species and even help to overcome the energy barriers of certain key reactions in the vicinity of $H_2$-forming sites.[10,11] Moreover, the ability to adsorb atoms and molecules depends primarily on the morphology of ASW: a highly porous ice can adsorb larger quantities of gas with respect to a compact one. The diffusivity of reactants and the access to reaction partners, as well as the sticking properties of atoms and molecules, are also greatly influenced by the surface morphology. Therefore, the morphology of amorphous water ice in the ISM appears to be the crucial parameter to thoroughly understand the energetics and dynamics of gas-grain reactions in molecular clouds.

[a] LERMA-LAMAp, UMR8112 du CNRS, Observatoire de Paris et Université de Cergy-Pontoise, 5 Mail Gay-Lussac, 95000 Cergy Pontoise Cedex, France. Tel: +33 1 3425 7078; E-mail: econgiu@u-cergy.fr
[b] DMFCI, Università di Catania, Viale Doria 6, 95125 Catania, Italy
† Present address: INAF, Osservatorio Astronomico di Capodimonte, Via Moiariello 16, 80131 Napoli, Italy

While there is quite a general consensus that interstellar water ice is mainly amorphous,[12] the nature of its morphology still remains poorly known. Ice porosity is identified in the laboratory through the small two-peak absorption feature (at 3720 and 3696 cm$^{-1}$) due to OH-dangling bonds of the porous ice structure. To our knowledge, there has been to date no detection of such a porosity signature in the infrared spectra of interstellar water ice, perhaps suggesting that it may have a compact nature.[13,14]

Laboratory simulations have indeed confirmed that interstellar porous ice analogues can be compacted quite efficiently by UV irradiation[15] and by cosmic ion bombardment[16,17] in a time comparable to the estimated ice mantle lifetime. Similarly, amorphous water ice exhibits a compact structure when it is formed by simulating water formation in space at low temperatures via surface reactions between $O_2$ and H-atoms.[18]

The present study is intended to provide another piece of evidence that porous ASW undergoes a compaction process also in consequence of surface reactions leading to molecular hydrogen formation. In particular, we show that a thin highly porous water ice film is gradually transformed into a more compact structure upon D-atom exposure. The loss of porosity appears to be due to the local temperature increase of the ice when two hydrogen atoms recombine on the icy surface. The release of $H_2$ formation energy in the vicinity of $H_2$-forming sites thus might change the local morphology, acting like a sort of local annealing. Such a process can therefore contribute to compacting amorphous water ice mantles in the ISM concurrently with the aforementioned processes.

The article is organized as follows. In Section 2, we provide a brief description of the experimental set-up and of the experimental procedures used to carry out the experiments. Section 3 is devoted to the experimental results: experimental data are shown, discussed and analysed. A simple model is used to simulate the evolution of the desorption curves subsequently to a degradation of the water ice morphology. In Section 4, we present the comparison between the model results obtained in Section 3 and another well-established model elaborated by our group and described elsewhere.[19] Section 4 also presents a discussion on the mechanisms responsible for the decrease of the ice porosity. Finally, in Section 5, our experimental results are discussed with respect to their important astrophysical implications.

## 2. Experimental apparatus and procedures

The experiments described in this paper are performed using the set-up FORMOLISM (FORmation of MOLecules in the InterStellar Medium) located at the astrophysical laboratory of the University of Cergy-Pontoise (France), within the teams of the LERMA laboratories of the Observatory of Paris. Only parts of the system which were used for the present work are briefly described below. Further details can be found elsewhere.[20]

The apparatus consists of an ultra-high vacuum stainless steel chamber, with base operating pressure lower than $10^{-10}$ mbar. At the centre of the main chamber, the sample holder (a copper cylinder block) is thermally connected to the cold finger of a closed-cycle He cryostat. The temperature is measured with a calibrated silicon diode clamped on the sample holder and controlled by computer to ± 0.2 K with an accuracy of ±1 K in the 8−400 K range. The porous-ASW samples are grown by admitting water vapour into the chamber through a leak valve: in this way, water vapour can diffuse into the chamber and freeze out on the copper substrate previously cooled to 10 K. This method, also known as "background deposition", allows to control the amount (and thickness) of the deposited highly porous ASW within the accuracy of the pressure gauge reading, i.e. ± 0.1 monolayer (1 ML = $10^{15}$ molecules cm$^{-2}$).

In order to avoid any kind of interaction with the copper substrate, the porous ice samples are grown on top of a 100-ML compact ice film previously grown using a calibrated microchannel array doser located at a distance of 2 cm from the copper surface maintained at 120 K.

Two triply differentially pumped atomic/molecular beam lines are connected to the main chamber and aimed directly at the surface of the sample. Each beam line is equipped with a micro-wave frequency cavity where molecular species (deuterium, in this work) are dissociated and then injected into the main chamber. A beam of hydrogen atoms has also been used in our experiments giving similar qualitative results. Nevertheless, the signal-to-noise ratio is lower in the case of hydrogen-atoms, so here we present only the results concerning the exposure of porous-ASW ice to D-atoms. The flux of the $D_2$ beam in the present experiment is $9 \times 10^{12}$ molecules cm$^{-2}$ s$^{-1}$, as experimentally estimated. The efficiency of the $D_2$ dissociation is about 70 %. Using the *King and Wells method*,[19] it has been evaluated that, during the irradiation phase, the sample is exposed to ~1.2 × $10^{13}$ D-atoms cm$^{-2}$ s$^{-1}$.

The sticking coefficient of $D_2$ (and of D) can change with the morphology of the ice. For these experiments, we determined that a 30-second dosing of $D_2$ on compact ice corresponds to ~ 0.11 ML of adsorbed $D_2$ and to ~ 0.17 ML of adsorbed $D_2$ in the case of a highly porous ice.

In the upper part of the main chamber, a quadrupole mass spectrometer (QMS) is installed on a rotatable flange and is employed for the detection of products entering the chamber or coming off the sample. During temperature-programmed desorption (TPD) experiments, the QMS head is located 3 mm in front of the sample in order to maximize the $D_2$ ion count signal of desorbing molecules. TPDs are performed using a linear heating ramp of 10 K/minute.

We have recently shown that TPD spectra of $D_2$ molecules, used as a probe gas, serve as a powerful tool to characterize the ice morphology.[21] In previous works, Hornekær et al.[22] measured a continuous broad distribution of binding energies of $D_2$ adsorbed on non-porous and porous ASW films, while Perets et al.[23] fitted the experimental TPD curves of $D_2$ using a rate model with a discrete set of adsorption energies. Other authors have used other simple and volatile molecules such as Ar, $N_2$, $O_2$, CO for characterizing the morphology water ice.[24,25] In the present work, we chose $D_2$ molecules to study the properties of the ASW ice samples. The advantage of using $D_2$ is twofold: i) this species is particularly weakly bound to the water ice surface and ii) desorption occurs below 30 K, hence before a significant transformation of the ice structure and porosity destruction by annealing can take place.[26] It is therefore possible to study the changes in morphology without thermally inducing them. Secondly, $D_2$ molecules can diffuse within the interconnected network of pori of ASW at 10 K, namely the temperature chosen for our experiments, and relevant to interstellar molecular clouds.

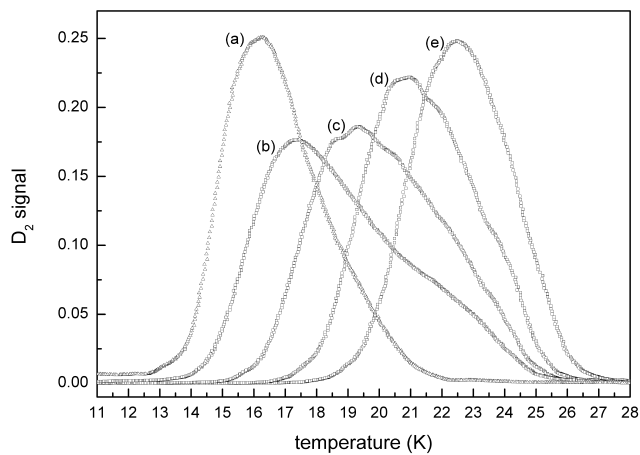
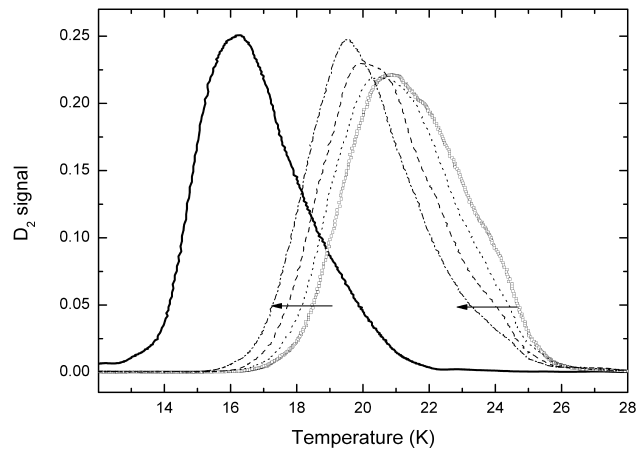

**Fig. 1** Comparison of TPD spectra obtained after 30 seconds of $D_2$ exposure on compact ice (a) and on highly porous ice films of different thickness (grown over a compact 100-ML ice substrate): compact + 1ML porous ASW (b), compact + 2ML porous ASW (c), compact + 4ML porous ASW (d) and compact + 8ML porous ASW (e).

**Fig. 2** Normalized TPD spectra of 0.11 ML of $D_2$ from compact ASW (solid line) and of 0.17 ML of $D_2$ from a 4-ML porous ASW film as deposited (open squares), after 64 min (dotted line), after 224 min (dashed line) and 288 min (dash-dot line) of D-atom irradiation. The arrows are meant to emphasize the shift of the TPD curves towards lower temperatures following D-atom irradiation.

Figure 1 shows a comparison between several TPD spectra of $D_2$ from a compact ASW (curve a) and from different thickness of highly porous ices grown over compact ice (curves b, c, d and e). Prior to each TPD experiment, the sample is annealed to 30 K in order to assure that no $D_2$ molecules are adsorbed on the ice surface and also to stabilize the surface morphology before subsequent heating-cooling runs between 10 and 30 K. These TPD spectra show clearly a high sensitivity to the ice surface roughness: their peaks shift towards higher temperatures as the porous ice network is progressively formed. $D_2$ molecules have completely desorbed by 21 K in the case of the compact non-porous substrate. When a layer of porous ice is deposited on the surface, $D_2$ is more bound to the ice surface and the desorption occurs at higher temperatures. In other words, increasing the porous ice thickness favours the formation of more energetic binding sites. In addition, as shown in Figure 1, the overlap between the $D_2$ TPD curves from compact ice and the ones from porous ice becomes gradually smaller with the increase of the porous ice thickness. Thus, $D_2$ TPD spectra prove to be very effective for probing water ice morphology and thickness.

## 3. Experimental results

TPD spectra of a fraction of a monolayer of $D_2$ molecules are performed before and after D-atoms exposure of the porous ASW film to probe its morphology and the possible loss of porosity as a function of D-atom fluence.

Figure 2 shows the effects of D-atom exposure on the morphology of a 4-ML porous ASW film. The solid line represents a $D_2$ desorption spectrum from compact ice, while open squares show the thermal $D_2$ desorption from a non-irradiated 4-ML porous ASW ice. The other TPD spectra of Figure 2 show $D_2$ desorption after subsequent D-atom bombardment of the porous ice sample kept at 10 K. Increasing the D-atom fluence, it is apparent a progressive shift of the desorption peaks towards lower temperatures, as the arrows indicate in the figure. Our set of experiments demonstrates that this effect is strictly due to D-atom irradiation of the ice. In fact, no changes of the ice morphology (namely, no changes of the TPD spectra) are observed in a porous ice film even after long exposures to molecular deuterium.

### Analysis by direct inversion of TPD curves

Each TPD spectrum, expressed in number of molecules coming off the surface per unit time, are known to follow the *Polanyi-Wigner* equation:[27]

$$dN/dt = r = -A\, v^n\, exp(-E_v/kT) \qquad (1)$$

where $r$ is is the rate of desorption, $A$ is the pre-exponential factor, $v$ is the number of molecules adsorbed, $n$ is the order of reaction, $E_v$ is the energy barrier for desorption, $k$ is the Boltzmann constant and $T$ is the absolute temperature of the surface.

By inverting Eq. (1), the desorption energy $E_v$ can be calculated as a function of coverage $v$ (number of molecules still adsorbed on the surface). For $D_2$ physisorbed on ASW ice the order of desorption is $n = 1$. If we take $A = 10^{13}$ s$^{-1}$ and assume that it is independent of $v$, $E_v$ can be calculated for each TPD spectrum as follows:[28]

$$E_v = -kT\, ln\,(r\,/\,A\,v) \qquad (2)$$

Figure 3 displays the $D_2$ coverage *vs* the desorption energy $E_v$ on a 1-ML porous ASW film as calculated using Eq. (2). These curves give us the number of $D_2$ molecules still bound to the ice sample as a function of their binding energy.

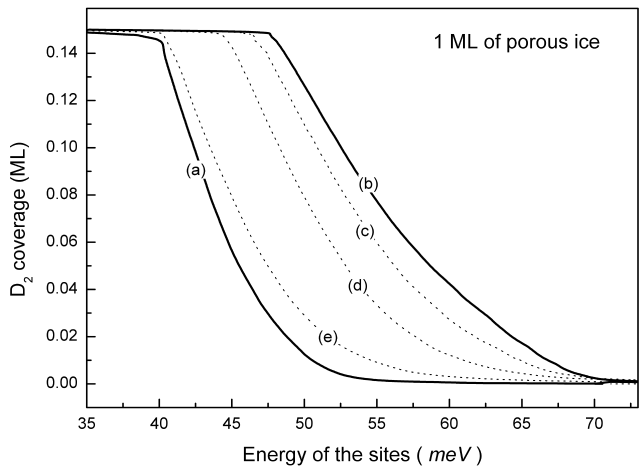

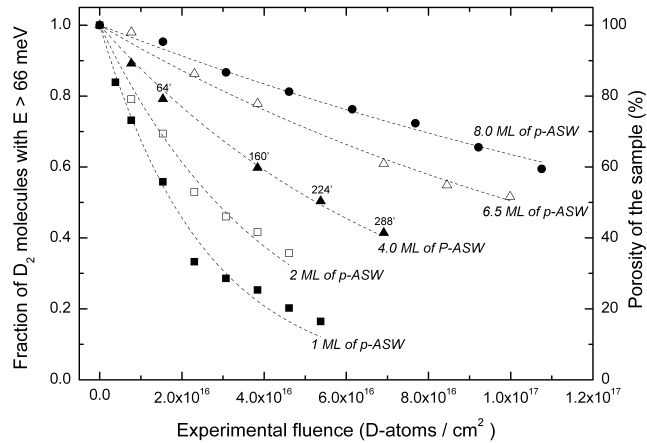

**Fig. 3** $D_2$ coverage *vs* desorption energy based on the inversion analysis of $D_2$ TPD curves from a compact ASW ice (a), and from a 1-ML porous ASW ice as deposited (b), after 64 min (c), 160 min (d) and 224 min (e) of D-atoms irradiation. All curves were smoothed for clarity using 20-datapoint Fast Fourier Transform (FFT) averages.

Growing an overlayer of porous ice increases not only the number of adsorption sites but also the energy distribution of binding sites associated to the ASW ice surface. This analysis emphasizes the existence of a small fraction of more energetic adsorption sites that are not present on compact ASW. We can see that binding sites with energy greater than 53 meV are certainly to be assigned to porous water ice. On the other hand, binding sites with lower energy may belong to either a compact or a porous water substrate. The binding sites below this energy range are those from which $D_2$ molecules desorb at a temperature below 21 K (see Fig. 1).

Via a statistical model for adsorption and desorption of $D_2$, it is possible, however, to disentangle between the two binding-energy distributions on compact and porous ice. In a previous work,[19] we succeeded in modelling the distribution of hydrogen molecules on their adsorption sites so that TPD spectra of various exposures and from compact or porous ASW can be simulated. As shown elsewhere,[21] this approach allows us to infer the energy distribution of our ASW ice sample as a combination of "compact" and "porous" binding sites. Of interest here is the degradation of porosity, so we shall focus on the binding sites that certainly are associated to a porous ice structure, i.e. sites of energy higher than 53 meV. Fig. 3 shows that for 1 ML the number of the most energetic binding sites decreases gradually with the increase of the D-atom fluence on the ice sample. We can see that the number of molecules with adsorption energy of 60 meV bound to a 1-ML porous ASW film decreases, with respect to the non-irradiated sample (curve b), by ~ 30% in the sample irradiated with 64 min of D-atoms (curve c), by ~ 70% in the sample irradiated for 160 min and by ~ 95% in the sample irradiated for 224 min of D-atoms. This confirms the progressive destruction of the "porous" binding sites, namely the compaction of the water ice sample after D-atoms exposure.

This experimental technique, together with the statistical model of adsorption and desorption of $D_2$, allows us to derive the binding-energy distribution of each porous ASW sample before and after irradiation, enabling to follow their evolution

**Fig. 4** Normalized fraction of $D_2$ molecules that can adsorb on the porous ice surface with an energy greater than 66 meV *vs* fluence of D-atoms. The fluence of D-atoms was used in this plot since it is a better parameter whenever an application in the astrophysical context is desirable. The D-atom fluence is proportional to the irradiation time according to the D-atom flux given in Section 2. Thus, data-points of the 4-ML curve, for example, can be directly compared with the curves in Figure 2 to have a quantitative estimate of the loss of porosity with time (exposure time in minutes is indicated above the 4-ML data-points).

as the surface roughness becomes smoother.

**Influence of the thickness of the ice**

The compaction of porous ASW ice due to D-atom irradiation was investigated for various thickness of the ice film. Figure 4 summarizes all the experiments performed in the present study, i.e., for porous ice thickness of 1 ML, 2 ML, 4 ML, 6.5 ML and 8 ML. In Fig. 4, we show the normalized fraction of $D_2$ molecules with adsorption energy greater than 66 meV (corresponding adsorption energy of molecules coming off at a surface temperature greater than 22.5 K). We quantified the porosity of each ASW ice sample as the surface area subtended by the TPD trace for temperatures greater than 22.5 K, since in this range the desorbing molecules were almost certainly adsorbed on "porous" binding sites. In fact, the trailing edges of the TPD curves give information on the distribution of adsorption sites available in the ice, providing thus a "signature" of the water ice surface roughness. We will discuss later the choice of the lower limit energy value of 66 meV.

We can see in Fig.4 that, for each ice thickness, the number of molecules that can adsorb on the surface with an energy larger than 66 meV decreases gradually as the D-atoms fluence is increased on the porous ice sample. It is clear that the experimental points of each ice thickness are well fitted by an exponential decrease, function with a unique adjustable parameter. Each exponential function is then identified by a characteristic value that expresses either the time of D-atom irradiation ($t_c$) or (to put it in other terms) the D-atom fluence ($\phi_c$). Either way, it represents the number of D-atoms impinging on the porous ice surface necessary to reduce by 64 % its initial porosity; the values of $t_c$ and $\phi_c$ for each exponential curve are reported in Table 1.

In Figure 5, the best-fit values of $\phi_c$ are plotted as a function of the porous ice initial thickness: for a thicker ice film a greater fluence is needed to obtain the same degree (in %) of reduction of porosity. This result was however expected as a thicker ice film implies a greater number of energetic binding sites to be destroyed. We can also notice that, within experimental errors, there is a linear correlation between the ice thickness and the characteristic fluence $\phi_c$. From the slope of the linear fit, it is possible to extract a characteristic fluence $\phi_0$ that gives us the hydrogen atom fluence needed to destroy the ice porosity of one layer of highly porous water ice. The value obtained is $\phi_0 = 2.2 \times 10^{16}$ D-atoms cm$^{-2}$ per H$_2$O ML.

**Table 1** Characteristic time $t_c$ and fluence $\phi_c$ of the exponential decay obtained by fitting the curves of Fig. 4.

| Porous ice thickness (ML) | $t_c$ (minutes) | $\phi_c$ (atoms cm$^{-2}$) |
|---|---|---|
| 0 | 0 | $0 \times 10^{16}$ |
| 1 | 125 ± 11 | ( 3.7 ± 0.7 ) × $10^{16}$ |
| 2 | 172 ± 5 | ( 4.1 ± 0.3 ) × $10^{16}$ |
| 4 | 269 ± 25 | ( 7.0 ± 0.7 ) × $10^{16}$ |
| 6.5 | 588 ± 20 | ( 1.4 ± 0.1 ) × $10^{17}$ |
| 8 | 766 ± 21 | ( 1.8 ± 0.1 ) × $10^{17}$ |

Now, we have to spend some words about the choice of the binding-site energy values greater than 66 meV for obtaining information on the ice morphology. A first natural choice could have been a lower boundary energy of 53 meV (corresponding energy of 21 K on TPD spectra and highest temperature at which we observe D$_2$ desorption from the compact ASW ice sample). For the case of the 8-ML porous ice film, however, the D$_2$ desorption occurs at temperatures too high that we could not get any information on the 53 meV energy sites from the TPD trailing edge. As seen in Fig.1, thermal desorption of D$_2$ from porous ice of thickness going from 1 to 8 ML shows that the overlap between all the TPD tails is confined within the range 22.5 – 25 K, thereby comes the choice of the energy lower limit of 66 meV.

It should be stressed, however, that any adsorption energy value chosen in the temperature range 22.5 – 25 K could have been used and still, both qualitatively and quantitatively, we would have obtained the same results. In fact, the inset of Fig.5 shows that different values of $t_c$ for the 1-ML porous ASW film are comprised in a very narrow set of values (to be considered identical within error bars) although they were calculated using different lower binding-energy boundaries. The same behaviour was observed for all the ice thickness studied. This means that the destruction of porosity seems to be independent of the binding energy of the sites, so that they are all destroyed by the same degree at once. This is not an obvious outcome though, because the deepest adsorption sites should, in principle, have a larger probability to be visited by the impinging D-atoms, thus have a larger probability to host a recombination reaction and be destroyed. Nonetheless, this is probably hindered by the ever-present molecular fraction of D$_2$ which is in competition with D-atoms in the pursuit of a free adsorption site. In fact, TPD spectra from D$_2$-saturated porous ice surface after various exposures to D-atoms confirm this hypothesis. D$_2$-saturated TPDs have traces which go from 10 K to 25 K so that the whole range of binding sites of energy between 35 and 70 meV is scanned. These spectra show – although they could not be utilised in the analysis for a reliable quantitative study – that the desorption rate is reduced along the whole spectrum of adsorption energies as the D irradiation time is increased. This piece of evidence as well indicates that binding sites of different energy depth are homogeneously eroded by recombination reactions between hydrogen atoms occurring on the surface of a highly porous water ice film.

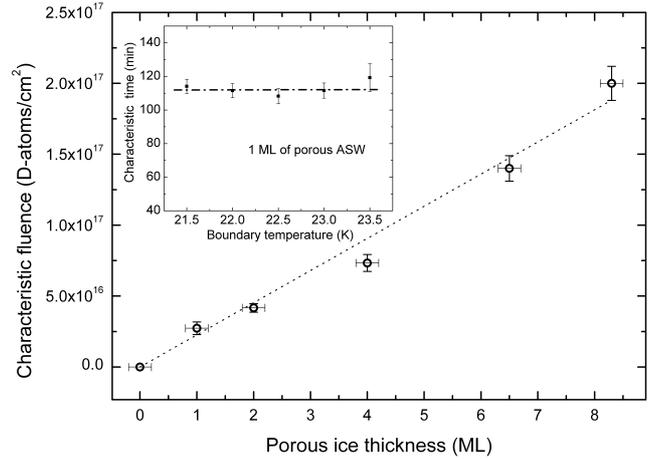

**Fig. 5** Characteristic fluence $\phi_c$ of the exponential decay *vs* porous ice thickness. Inset: characteristic time $t_c$ as a function of the energy lower limit used for evaluating the loss of porosity (*see text for details*).

## 4.Discussion

### Modelling of TPD spectra and validity of the direct inversion analysis.

In the present work we have employed the *inversion analysis method* of TPD spectra to obtain the distribution of the binding-site energy on the surface of each porous ASW ice samples. This method proved very sensitive to the evolution of the surface morphology and also allowed us to provide quantitative results of the rate at which the porosity of the ice film is destroyed by D-atom irradiation. To complement this pure experimental technique, we also modelled the D$_2$ TPD spectra from porous water ice using a more sophisticated model[19] by which we derived the binding-energy distribution of each porous ice sample. This model uses a simple statistical approach to simulate the distribution of D$_2$ on the adsorption sites and how surface temperature and coverage affect it. There are two fundamental assumptions in this model: 1) the molecular hydrogen is continuously in complete thermal equilibrium with the surface, even during the heating process; 2) the molecular hydrogen explores all the available adsorption sites. In this model, the binding-energy distribution is described by a polynomial function $g(E)=a(E_0 -E)^b$, where $a$, $E_0$ and $b$ are parameters to be determined from the TPD analysis, i.e., the fit of experimental TPD spectra. At a given temperature and a given coverage, this distribution is assumed to be populated following a Fermi-Dirac statistical law, because one adsorption site can be filled by one molecule solely.

Figure 6 shows some synthetic D$_2$ TPD curves obtained with the statistical model and how they reproduce well the experimental $D_2$ TPD spectra. The fit of curve (b) corresponds to the desorption of $D_2$ from the 8-ML porous ice as deposited, i.e., before being exposed to D-atoms. The free parameters have the same values as in a previous work,[19] except the value of $E_0$ that was fixed to 81 meV instead of 78.9 meV. This is fully compatible with a difference of temperature accuracy of 2 % (0.4 K at 20 K), that could occur between two sets of experiments. The quality of the fit is as good as in the results obtained previously.[19]

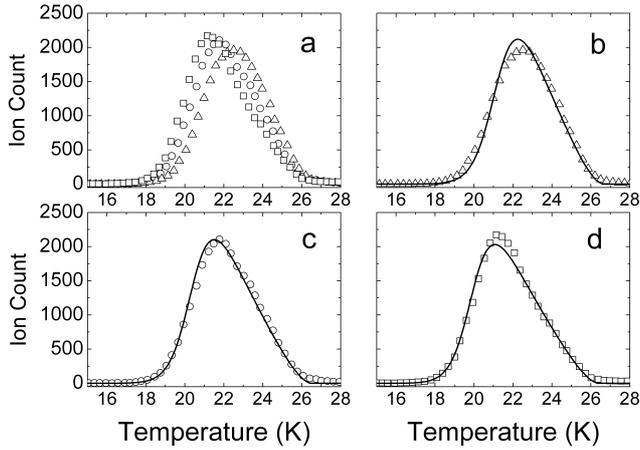

**Fig. 6** Scatter points: TPD curves of 0.17 ML of $D_2$ desorbing from an 8-ML porous ice film: as deposited (triangles in a and b), after 256 min of D exposure (circles in a and c) and after 448 min of D exposure (squares in a and d). Solid black lines: TPD simulations (b,c,d) obtained using a statistical model of $D_2$ adsorption and desorption from ASW ice.

The curves (c) and (d) of Fig.6 are fits from a 8-ML porous ASW sample irradiated with 256 and 448 minutes of D-atoms respectively. The experimental data were simulated by scaling down the number of accessible sites on the ice with irradiation time. This simple strategy was adopted on the basis of the experimental results discussed in the previous section: the destruction of the binding sites on a porous ice film after D-atom exposure seems to be independent of their binding energy depth. Therefore, subsequently to each run of atomic hydrogen exposure, the whole distribution of adsorption sites of the porous ice film is simply scaled down by a constant factor $R_p$. The gradual reduction of the porosity – that can be described now by $R_p$ – can also be plotted versus time of D-atom fluence. Likewise, an exponential decrease is observed also in this case. The characteristic time $t_c$ found here is slightly higher (by ~ 15 %) than the value determined for the same ice thickness through the exponential decay shown in Fig.4. This small difference may be due to the contribution of the leading edges in the TPD profiles which are taken into account in the Fermi-Dirac equilibrium model, whereas in the classical inversion method, only the tails of the TPD traces contribute to the evaluation of the energy distribution of the adsorption sites.

Given the small difference between the results obtained using a sophisticated model and the simple inversion of TPD spectra, we can fairly assume that the *direct inversion analysis* is a simpler and still a capable tool for correctly determining the reduction of the ice porosity with only one adjustable parameter ($t_c$).

**Origin of the decrease of the porosity**

The destruction of the disordered and fluffy structure of a highly porous ASW ice film is evidently induced by D-atoms irradiation as this work has demonstrated. The origin of the decrease of the ice porosity following exposure to D-atoms lies in the formation energy (4.48 eV) released when the adsorbed atoms react to form $D_2$ molecules. Since a fraction of the formation energy is deposited in the ice,[9,29] this exothermic reaction occurring at the surface and within the pori of the ice causes local heating around the site of the reaction, acting like a local annealing. As we stated earlier, the rate at which the porosity is reduced does not depend on the binding energy of each adsorption site which is destroyed. This means that the probability that each adsorption site is occupied by a D-atom is equal. This, in turn, implies that the diffusion barriers between the sites are the same. Indeed, this may also mean that the hopping barrier between two sites is not linked to the binding energy of a specific site, perhaps owing to the very disordered nature of highly porous water ice. Another scenario entails the possibility that co-adsorbed $D_2$ molecules affect the distribution of D-atoms on the surface, in the same way that isotopes[30] or even ortho- and para-states of the same molecule[31] are in competition for the same adsorption sites. This contention could reduce the importance of the binding energy of the sites as far as the residence time of atoms is concerned, because molecules have also a greater chance to occupy the most binding sites.

Extensive experimental studies demonstrate that porous ASW films undergo an irreversible densification upon annealing,[1] and that the destruction of the porous network is complete by 110 K.[32,33] The re-arrangement of water molecules during densification by annealing implies the diffusion of $H_2O$ molecules on the surface. In the specific case of interest here, that is porous ice films kept at 10 K, it would be perhaps more appropriate to speak of restricted local migration rather than diffusion. In fact, the water molecules are frozen in position and, owing to the local heating produced by nascent $D_2$ molecules, can only undergo minimal displacements or rotational motions. Let's therefore estimate an activation energy for this diffusion, necessary to rearrange water molecules in the vicinity of an adsorption site. Jung *et al.*[34] measured an activation energy for self-diffusion of surface molecules in compact water ice of 14 kJ mol$^{-1}$ (146 meV). It is important to note, however, that this value can be fairly considered as an upper limit for highly porous ice films. In fact, surface molecules will be highly disordered and this will result in a smaller diffusion energy barrier.[34,35] To estimate the activation energy for diffusion $E_d$, we assume that one reaction (D + D → $D_2$) will provide locally a temperature T that gives the probability of diffusion close to one: $P_d = 1 = \alpha \exp(-E_d/RT)$, with $\alpha = 10^{15}$ being the molecular dynamics time scale, value reasonably chosen according to other authors.[36] We can find now a set of appropriate couples T and $E_d$ that satisfy the condition $P_d = 1$, namely [38 K, 110 meV], [90 K, 270 meV] and [140 K, 415 meV]. These temperatures were chosen as known thresholds linked to the thermal annealing of porous water ice: 38 K represents the beginning of the irreversible change of high density into low density ice,[1] 90 K is the low boundary temperature required to grow compact ice[32] and 140 K is the highest temperature before crystallization occurs.[33] As for the values of $E_d$, 110 meV, 270 meV and 415 meV, they represent respectively 2.5%, 6% and 9% of the total energy released upon formation of one hydrogen/deuterium molecule.

In one of our recent studies, we estimated that the fraction of the formation energy released into the ice is comprised between 40 and 60%.[9] Because this energy budget is deposited in one adsorption site where the number of water molecules bonded with the adsorbate in the potential minimum is ~3.3,[37] and that one surface water molecule is in turn coordinated with 2 or 3 other water molecules,[38] the mean energy transferred to a single water molecule is 5 to 7% of the $H_2$ formation energy. Therefore, allowing for the local heating to cause only the annealing of the ice, we can predict that not every reaction leading to a $D_2$ molecule is able to contribute to the destruction of the ice morphology. In fact, the released energy per molecule is not well above the activation energy for the diffusion of water molecules (9%).

We try now to estimate the probability of destroying one adsorption site per each atomic recombination. It has already been evaluated (Fig.5) that the dose required to decrease the porosity by 64% of one layer of p-ASW is a fluence $\phi_0 = 2.2 \times 10^{16}$ D-atoms cm$^{-2}$ per $H_2O$ ML. Although, the recombination efficiency $R$ is not one on porous ASW: it was found to be ~0.3[34] and ~0.1[6]. So, if we take a rough arithmetic mean of previous measurements of $R$ (= 0.2), the number of reactions needed to decrease the porosity by 64% is thus $\phi_0 \times R/2$ = 2.2×10$^{15}$ reactions cm$^{-2}$ per $H_2O$ ML (dividing by 2 is required because two atoms form one $D_2$ molecule). The number of required recombinations evaluated conservatively in this way has to be considered as an upper limit, because multiple annealing events occur inside pori during the migration of every nascent very energetic (they possess initially 95% of the 4.48-eV energy budget) $H_2$ molecule.[39] This is certainly true in thicker ice films where the porous structure is fully developed, but it is already evident for thin ASW films as well. A rough quantitative estimate of the densification effect provides an average number of annealing events (all energetically similar to the first one) per recombination that increases from 1 for thin (1-3 monolayers) mantles to perhaps 5 for ASW ice films thicker than 20 or 30 ML. In fact, previous experimental results suggest that many more than 5 collisions within the pori are needed before accommodating newly formed $H_2$ molecules in the ice.[8,39]

The complex structure and surface geometry of the porous ice film (containing crevices, tunnels and micropori) leads to a large variety of adsorption sites, namely, to a wide energy distribution. Studying the adsorption of $D_2$ on 10-ML film of porous ice,[19] we estimated $3 \times 10^{15}$ cm$^{-2}$ (3 ML) to be the number of adsorbing sites on $10^{16}$ cm$^{-2}$ (10 ML) $H_2O$ molecules. In the light of this result, the number of adsorbing sites in our porous ASW sample is ~ $0.3 \times 10^{15}$ per $H_2O$ ML. Therefore, taking into account both the density of adsorption sites per ML of ice and the number of atomic recombination responsible for the compaction of the ice, we can conclude that only one reaction out of 7 contributes to the compaction of the ice structure. Possible causes of the inefficiency of this process include the following: i) not all the reactions provide enough energy; ii) not all the adsorption sites can be easily destroyed/re-arranged; iii) the fraction of energy deposited in the ice can be spread over more than one water molecule, or even exchanged with other co-adsorbate molecules.[9]

## 5. Astrophysical implications

In this last section, we want to assess the time necessary to achieve a considerable destruction of porosity as it applies to an interstellar dense cloud. Actually, we want to know whether or not the decrease of the ice porosity by atomic hydrogen exposure is astrophysically relevant.

The icy mantles covering dust grains in dark clouds are believed to be 40-100 molecular layers thick.[14] There is, however, already evidence that support that the ice cannot be fully porous. For instance, we have recently observed that the morphology of a water ice layer formed through the pathway $D + O_2$ on an interstellar ice analogue is compact. Indeed, this result was also confirmed by a recent article,[18] even though our experiments are performed under conditions that are more relevant to astrophysical scenarios.

By the present study, we can say that, if the growth of the water ice is 1/14 lower than the H accretion, the water layer porosity is reduced as fast as the icy mantle grows. This possibility should happen as soon as the H flux is 14 times more intense than the O flux. This is the case in most interstellar environments with number densities of $10^3$ cm$^{-3}$ or less where hydrogen is in the form of H-atoms and water formation on dust grains proceeds efficiently via the $H + OH$ pathway[40]. On the other hand, in the densest and coldest molecular clouds hydrogen is in the form of $H_2$ and O-atom is the most abundant species in atomic form: for a cloud density of $10^5$ cm$^{-3}$, the ratio H/O is estimated to be ~ 1/7.[41] As for typical dark clouds with number densities between $5 \times 10^3$ and $5 \times 10^4$ cm$^{-3}$, $H_2$ is again the dominant species in gas phase and participates in surface reactions leading to water molecules.[40]

The interstellar flux of hydrogen atoms impinging on dust grains can be calculated as follows:

$$\Phi_H = \frac{1}{4} n_H v_H \qquad (3)$$

where $n_H$ is the H-atom density in gas phase and $v_H = (8kT/\pi m)^{1/2}$ is the mean velocity of an hydrogen atom. Given a typical molecular cloud with $n_H = 2.3$ cm$^{-3}$ and T = 10 K,[7] the estimated H flux is $8.35 \times 10^{11}$ atoms cm$^{-2}$ yr$^{-1}$.

In Figure 7 we present the evolution of a porous film of water ice in a typical molecular cloud. The dashed line represents the icy mantle thickness following a mean formation rate of 100 ML in $10^6$ years.[36] The solid line shows the evolution of the thickness of the porous layer as it is concurrently exposed to the mean H-atom flux as calculated above and undergoes a porosity destruction rate calculated using the characteristic fluence $\phi_0$ found in this work. Fig.7 shows that, due to H-atom irradiation, the porous film never exceeds a thickness of 55 ML and it is completely destroyed within $3 \times 10^6$ years. Nevertheless, this length of time ought to be considered as an upper limit and a rather conservative estimate since the mean formation rate of water ice employed is based on two very optimistic hypotheses: 1) a very high efficiency of conversion to water by surface reactions and 2) water ice just formed is assumed to be porous. Moreover, should the H-atom flux be a factor of 2 higher or the actual formation rate be a factor of two slower, the thickness of the porous layer would never exceed 10 ML and would be completely compacted in less than $10^6$ years. Therefore, the destruction of porosity is completed in a lapse of time much shorter than the estimated lifetime of a molecular cloud ($3\times10^7 - 5\times10^8$ years).[16]

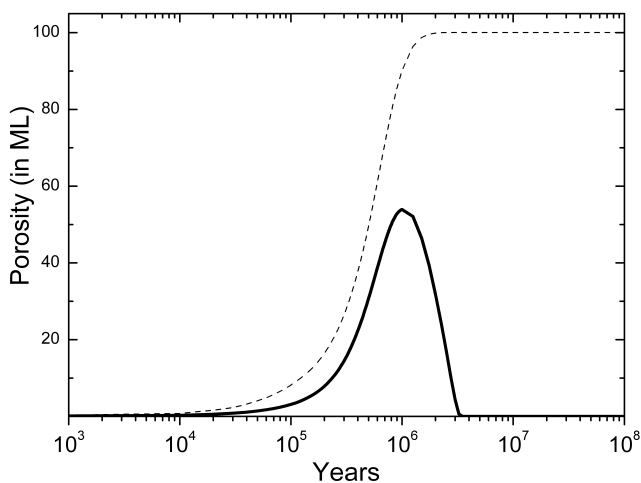

Fig. 7 Evolution of the thickness of a highly porous ASW mantle following grain surface reactions only (dashed line) and with concurrent H-atom exposure (solid line) in a typical molecular cloud (*see text for details*).

There are also other sources of compaction of water ice in dark clouds such as energetic ions,[16] the so-called cosmic rays. Nevertheless, as the penetration of the cosmic rays is large compared to the typical interstellar ice mantle depth, this effect is independent of the ice thickness. Thus, if the initial thickness of the porous film is not larger than a few tens of ML, the compaction due to atomic hydrogen irradiation is more efficient than the same effect due to the interaction with cosmic rays, which remains a rather rare event.

To conclude, we can say that the compaction of porous water ice as the consequence of H-atoms recombination should be a very efficient process in typical interstellar dark clouds. Hence, on the basis of our and other laboratory results, interstellar water ice undergoes a compaction process that is the result of H-atom irradiation, cosmic rays[16] and UV bombardment.[15] These experimental evidences lead us to suggest that water ice in space is almost certainly amorphous and non-porous (compact).

## Acknowledgements

We acknowledge the support of the French National PCMI programme funded by the CNRS, as well as the strong financial support from the Conseil Régional d'Ile de France through SESAME programmes E1315 and I-07-597R, the Conseil Général du Val d'Oise and the Agence Nationale de la Recherche.